# Epitaxial Sr-doped nickelate perovskite thin films and Ruddlesden-Popper phases grown by magnetron sputtering

Changhwan Kim[a,†], Min Young Jung[a,†], Yeong Gwang Khim [b,c,†], Kyeong Jun Lee[a], Young Jun Chang[b,c], and Seo Hyoung Chang[a]*

[a]*Department of Physics, Chung-Ang University, Seoul 60974, Republic of Korea*
[b]*Department of Physics, University of Seoul, Seoul 02504, Republic of Korea*
[c]*Department of Smart Cities, University of Seoul, Seoul 02504, Republic of Korea*

Sr-doped nickelate, $Nd_{1-x}Sr_xNiO_3$ (NSNO), perovskite thin films and Ruddlesden-Popper (RP) phases are actively investigated because of their physical properties, such as the metal-insulator transition and superconductivity. However, achieving epitaxial growth of NSNO perovskite and RP phase films in a sputtering system is challenging compared to pulsed laser deposition and molecular beam epitaxy, due to the difficulty in stabilizing nickel oxidation states and minimizing structural defects. Here, we used an off-axis radio frequency (RF) magnetron sputtering to fabricate epitaxial NSNO perovskite and RP phase thin films on $SrTiO_3$ (001) substrates, systematically controlling the growth temperatures. We investigated the thermal stability of the perovskite phase and the structural and electronic characteristics of the RP phase films. These findings provide valuable insights into the synthesis of nickelate RP phase films using RF magnetron sputtering, paving the way for scalable thin films fabrication technologies.

e-mail: cshyoung@cau.ac.kr

# 1. Introduction

Perovskite rare-earth nickelates, RNiO$_3$, have been intensively investigated due to their intriguing and tunable physical properties, such as metal-insulator transition (MIT), rich magnetic states, and charge disproportionation [1–3]. For instance, MIT transition temperatures, $T_{MI}$, can be modulated by factors such as the ionic radius of rare-earth elements, the electron/hole doping, and the thickness of nickelate thin films [4–7]. Moreover, superconductivity has been observed in infinite-layer nickelates, Nd$_{0.8}$Sr$_{0.2}$NiO$_2$, (the 112 phase) and Ruddlesden-Popper (RP) nickelates under high pressure conditions [8–11]. Despite these advancements, the high-quality epitaxial synthesis of perovskite and RP phase films has primarily been limited to specific growth techniques, such as pulsed laser deposition (PLD) and molecular beam epitaxy (MBE) [12–15].

Compared to other growth techniques, sputtering offers several advantages, including scalability for large-scale growth, cost-effectiveness, and a typically high deposition rate. Epitaxial growth of Nd$_{0.8}$Sr$_{0.2}$NiO$_3$ (NSNO) films grown on SrTiO$_3$ (STO) substrates has been extensively explored as a precursor phase for infinite-layer nickelates [8,12]. For systematic studies into the chemical reactions of the infinite-layer nickelates, achieving uniform and large-scale growth of NSNO films via sputtering is particularly effective. However, forming NSNO perovskite phase using sputtering presents significant challenges due to the nickel oxidation states exceeding 3+, which arise from Sr doping [16].

To achieve Sr-doped nickelate perovskite films via sputtering, it is crucial to address phase instability and to reduce defects induced by the plasma's high kinetic energy and thermodynamic effects. For instance, as temperature increases, bulk perovskite NdNiO$_3$ can decompose into NiO and RP phases, such as Nd$_2$NiO$_4$ and Nd$_4$Ni$_3$O$_{10}$ [17]. The off-axis radio frequency (RF) magnetron sputtering system effectively reduce the kinetic energy of plasma at the substrate while enabling precise control over the growth rate. Consequently, this approach enhances the quality of epitaxial growth, making it an optimal technique for this study.

Here, we present the stabilization of epitaxial Nd$_{1-x}$Sr$_x$NiO$_3$ (NSNO) thin films on SrTiO$_3$ (STO) (001) substrates using an off-axis RF magnetron sputtering system while controlling the growth temperatures. We investigated the physical properties of epitaxial NSNO perovskite and RP phases using X-ray diffraction, X-ray photoemission spectroscopy, and transport measurements. We observed growth temperature-dependent phase formation of NSNO, closely linked to the modulation of oxidation states of nickel. This study provides insights into the epitaxial growth of nickelate thin films using sputtering systems.

## 2. Experiment

NSNO thin films were grown on (001)-oriented STO substrates using an off-axis RF sputtering system with growth temperatures controlled from 520 °C to 670 °C. The STO substrates were cleaned and ultrasonically washed with Isopropyl alcohol (IPA), acetone, and deionized (DI) water. The target (Toshima Co. Ltd.) was polycrystalline and consisted of two phases, (Nd$_{0.8}$Sr$_{0.2}$)$_2$NiO$_4$ and NiO, designed to achieve a Nd/Sr to Ni stoichiometry of 1:1. An off-axis gun, with the target mounted and an RF power of 50 W, was used to control the growth rate at a slower pace compared to a typical growth rate of an on-axis gun.

Reciprocal space mapping (RSM) and 2θ/ω scans were performed using a high-resolution X-ray diffraction (HR-XRD, Bruker AXS D8) with monochromatic Cu K$_\alpha$ beam source (λ = 1.5406 Å). For detailed structural analysis, HR-XRD measurements were performed at the 3A and 5A beamlines at Pohang Accelerator Laboratory (PAL), South Korea [18]. X-ray photoelectron spectroscopy (XPS) was performed using an instrument (Thermo Fisher Scientific, Co.) equipped with an Al K$_\alpha$ X-ray source (1486.6 eV) and a micro-focused monochromator with a 400 μm spot size. Binding energy of Carbon 1s of 284.80 eV was used for calibration. Temperature-dependent resistivity measurements were carried out with a physical property measurement system (PPMS, Quantum Design Co.) using both the Van der Pauw method and a four-probe geometry with indium wire-bonded contacts.

## 3. Results and discussion.

Figure 1(a) shows a phase instability of NSNO thin films related to the possible decomposition at growth temperatures. Bulk rare-earth nickelates can decompose from a perovskite to the Ruddlesden-Popper (RP) phases and NiO. For instance, it was reported that a bulk perovskite $LaNiO_3$ was stable at the temperature lower than 1273 K under an oxygen pressure of about 0.2 bar [19]. However, as the temperature increased certain critical temperatures from 1273 K to 1523 K, the film was decomposed into $La_4Ni_3O_{10}$, $La_3Ni_2O_7$, and $La_2NiO_4$ RP phases with NiO and oxygen [19, 20]. The thermodynamic phase stability of bulk rare-earth nickelates including R = La, Sm, and Nd, has been experimentally investigated and theoretically understood in terms of Gibbs energy based on the enthalpy of formation, entropy, and heat capacity of the RP phases [19–21]. The bulk La-Ni-O phase diagram as functions of temperature and oxygen pressure provides the insight that the instability of NSNO perovskite phase and the phase boundary can appear down to 600 – 800°C under our oxygen partial pressure of the sputtering condition, e.g., $1.6 \times 10^{-5}$ bar, as shown in Fig. 1(a).

As shown in Fig. 1(b), we investigated the crystal structures of NSNO thin films grown on STO (001) substrates, denoted by NSNO1 to NSNO4, with controlling the growth temperatures from 520 °C to 670 °C, respectively, using HR-XRD. During the sputtering, total pressure was fixed at 50 mTorr with Ar and $O_2$ gas flows in ratio of 3:1, respectively. When NSNO films were deposited at 670 °C, both $(Nd_{1-x}Sr_x)_2NiO_4$ (NSNO-214) RP phase and a NiO phase were observed, as displayed by the NSNO4 (red line) of HR-XRD pattern in Fig. 1(b). This growth temperature of NSNO-214 RP phase was quite lower than the critical temperature of bulk RP phase synthesis but seems to be followed by our prediction based on the phase diagram of rare-earth nickelates. Note that the NSNO-214 phase exhibited highly preferred orientation with the out-of-plane direction of the substrate indicating of the existence of NSNO-214 (004) and higher reflections, e.g., the NSNO-214 phase (006), and (008), in Fig. 1(b). Based on the 2θ values of RP phase peaks, the out-of-lane lattice parameter of the NSNO-214 phase was about 12.13 Å, which was consistent with the reported value of bulk $Nd_2NiO_4$, 12.11 Å [22]. The NSNO3 film grown at 620 °C exhibited the NSNO-214 phase, a NiO, and an additional small peak, denoted by the magenta-colored inverse triangle in Fig. 1(b). The shoulder-like peak near the STO (002) can be assigned to $(Nd_{0.8}Sr_{0.2})_4Ni_3O_{10}$ phases [12, 23].

As the growth temperature decreased to 560 °C and 520 °C, we were able to synthesize the perovskite nickelate thin film denoted by NSNO2 and NSNO1, respectively, in Fig. 1(b). In the HR-XRD data, intriguingly, there were little specular reflections related to the RP phases and the NiO phase, which were formed in the films grown at higher temperatures (NSNO3 and NSNO4). The NSNO2 film exhibited both perovskite NSNO (001) and (002) Bragg reflections. Specifically, the existence of NSNO (001) reflection is the indicator to determine the well-ordered perovskite phase in the films. Compared with the NSNO1, the NSNO2 film has less oxygen vacancies concentration and better crystallinity because of smaller out-of-plane lattice parameter and the strong NSNO (001) peak, respectively. This well-defined surface was confirmed by the clear Kiessig (thinkness) fringes at around the NSNO (002) peak.

To study the detailed crystal structure of perovskite NSNO films, we measured reciprocal space mapping of NSNO1 and NSNO2 films at around STO (103) and (013) off-specular reflections, as shown in Fig. 2. Both NSNO1 and NSNO2 samples were epitaxially grown on the STO substrates with fully coherent and tensile strained. Figure 2 clearly showed the tetragonal symmetry of both NSNO1 and NSNO2 films. The L value of the NSNO1 (NSNO2) from the RSM near STO (103) plane was a similar value with that of the NSNO1 (NSNO2) from the RSM near STO (013) plane within the error-bar, 0.05%. The out-of-plane lattice parameter of NSNO2 sample was 3.79 Å whereas NSNO1 showed larger out-of-plane lattice parameter of 3.81 Å, well-matched with the HR-XRD data in Fig. 1(b). Note that the crystal structure of bulk $NdNiO_3$ and $Nd_{0.8}Sr_{0.2}NiO_3$ are orthorhombic structures with *Pbnm* space group [24]. Because of larger ionic radius of divalent Sr, the introduction of 10 – 20% doping induced the volume expansion of bulk $NdNiO_3$ compound up to 1.4 – 2.6%, respectively [24, 25]. Compared with the reported volume of bulk NSNO compound (220.754 Å$^3$), the NSNO2 (NSNO1) sample exhibited the volume expansion of about 4.7 % (5.2 %), which can be related to the effect of cation stoichiometry and oxygen vacancies [25].

To get further insight into the transport properties of perovskite and RP phases, we performed temperature dependent resistivity, $\rho(T)$, measurements of NSNO films, as shown in Fig. 3. We were able to classify their $\rho(T)$ curves into three groups: (1) NSNO1, (2) NSNO2, and (3) NSNO3 and NSNO4. As shown in Fig. 3(a), the NSNO1 sample showed an insulating behavior with very weak temperature dependence. Compared with the NSNO1 sample, the NSNO2 clearly exhibited a metallic behavior above 90 K and an insulating phase at lower temperatures. The onset temperature of metal-insulator transition (MIT), $T_{MI}$, was approximately 90 K. From [25], the bulk NdNiO$_3$ exhibited the MIT and with the increase of Sr/(Nd+Sr) ratio from 0 to 0.2 (hole-doping), the values of $T_{MI}$ were changed from 190 K to 108 K, respectively. The suppression of MIT by Sr doping were also understood by the relation between the $T_{MI}$ and the Sr doping from [6]. Furthermore, tensile strained NdNiO$_3$ films grown on STO (001) substrates can have lower values of $T_{MI}$ compared to those of bulk nickelates [26-28]. This transport result of NSNO2 was quite consistent with the characteristic of Sr-doped nickelate perovskite phases, which might be closely related to the Ni-O-Ni bond angle and connected with the magnetic phase transition from paramagnets to antiferromagnets [29].

Figure 3(b) shows insulating behaviors of both NSNO3 and NSNO4 samples consisting of the NSNO-214 and NiO phases. Compared with NSNO1, the resistivity of NSNO4 (NSNO3) sample was increased rapidly more than an order of magnitude as the decrease of temperature from room temperature to 200 K (150 K) and met the limit of the measurement. The insulating properties of NSNO samples grown at higher temperature than 620 °C originated from the instability of perovskite phase and were closely related to the formation of the RP and NiO phases, confirmed by structural analyses. To address the conduction mechanism for insulating NSNO films in Fig. 3(c), we plotted the ρ(T) curves of both NSNO3 and NSNO4 samples with the Arrhenius equation, as following

$$\sigma(T) = \sigma_0 \exp\left[-\left(\frac{E_a}{k_\mathrm{B} T}\right)\right] \quad (1)$$

where $E_a$ and $k_\mathrm{B}$ are an activation energy and Boltzmann constant, respectively [30]. Rather than the Mott variable range hopping model, the ρ(T) curves followed the activated conduction well. Using Eq. (1), the $E_a$ values of NSNO3 and NSNO4 samples were about 79 meV and 131 meV, respectively. The values were higher than the reported activation energy of RP phases (~ 20 meV) and smaller than that of NiO phase (~ 180 meV) [31, 32]. From the charge neutrality, the NSNO films had lower valence states close to Ni$^{2+}$, which can induce the localized carrier and exhibit higher resistivities.

To get further insight into the cation stoichiometries and valence states of NSNO samples, we measured the core level XPS spectra of Nd $3d_{5/2}$ and Ni $2p_{3/2}$ for four NSNO films, as shown in Fig. 4. As shown in Fig. 4(a), the main $Nd^{3+}$ peak near 981.8 eV exhibited a similar shape and binding energy. position across all four NSNO samples. The shake-down satellite peak adjacent to the main $Nd^{3+}$ peak could originate from a charge transfer between the oxygen ligand and the transition metal [33]. Additionally, the satellite peak at a binding energy of approximately 972 eV corresponds to the O K-$L_1$ edge, representing oxygen electrons transitions from the K-shell to the $L_1$ subshell [33]. Figure 4(b) shows the Ni $2p_{3/2}$ spectra of the four NSNO samples with their fitting curves. The fitting faced certain limitations due to the negative charge transfer nature of rare-earth nickelates and contributions from electron-coupling screening effects [34]. In NSNO1, the hump peak at 853.4 eV (denoted by the red dashed line) corresponds to the binding energy of $Ni^{2+}$. The peaks associated with $Ni^{3+}$ and $Ni^{4+}$ were 855.0 eV and 858 eV, respectively, aligning well with previously reported values [35, 36].

The Ni $2p_{3/2}$ peaks of samples grown at lower temperatures (NSNO1 and NSNO2) showed noticeably different $Ni^{2+}/Ni^{3+}$ ratios compared to those grown at higher temperatures (NSNO3 and NSNO4). Specifically, NSNO4 sample exhibited an enhanced intensity of the $Ni^{2+}$ hump compared to the other samples, with a slightly higher $Ni^{2+}$ binding energy of than NSNO1. This observation suggests the presence of defective nickel oxides, as confirmed by the XRD data in Fig. 1(b). The $Ni^{3+}$ content increases steadily as the growth temperature decreases, becoming more dominant relative to the $Ni^{2+}$. Correspondingly, the $Ni^{2+}/Ni^{3+}$ ratio as a function of temperature confirms the phase boundary between NSNO2 and NSNO3, as discussed in Fig. 1.

A potential source of phase instability in nickelate thin films grown by RF magnetron sputtering is the formation of $Ni^{2+}$ phases. To mitigate defective nickel oxides and achieve NSNO perovskite thin films, low temperatures and high oxygen partial pressures are required, However, temperature lower than the optimal conditions hinder the crystallinity of the perovskite phase. To overcome this limitation, the off-axis geometry in RF magnetron sputtering allows precise control over the growth rate of NSNO films, enabling a delicate balance necessary for adatom migration at the surface and the formation of the perovskite phase while suppressing the formation of $Ni^{2+}$ phase. By integrating off-axis RF magnetron sputtering with phase diagram studies, this work provides valuable insight into achieving high quality nickelate perovskite and Ruddlesden-Popper phase thin films. Incorporating recently developed machine-learning approaches could further optimize these processes, opening the door to scalable and automated thin film fabrication technologies [37, 38].

## 4. Conclusions

In summary, we fabricated epitaxial nickelate perovskite thin films grown on SrTiO$_3$ (STO) (001) substrate by using RF magnetron sputtering. By systematically controlling the growth temperature, we achieved the perovskite Nd$_{1-x}$Sr$_x$NiO$_3$ (NSNO) phase at 520 °C and 560 °C, as confirmed by HR-XRD. At higher growth temperatures, the formation of the NSNO-214 Ruddlesden-Popper (RP) phase and NiO was observed, attributed to the thermodynamic instability of Ni$^{3+}$ state. This was consistent with the increase of Ni$^{2+}$ peak and resistivity from the XPS spectra and temperature dependent resistivity, respectively. Our work provide a comprehensive understanding of the phase boundaries for the perovskite NSNO thin films and offer valuable insight into the synthesis of high-quality nickelate thin films.

**Acknowledgments**

The work was supported by the Chung-Ang University Graduate Research Scholarship in 2021 (awarded to C.K.) and was supported by Basic Science Research Programs through the National Research Foundation of Korea (Grants No. NRF-RS-2023-00252772, RS-2023-00220471).

**Figure Captions**

**Figure 1** | (a) Schematic of phase segregation of $Nd_{1-x}Sr_xNiO_3$ (NSNO) perovskite as a function of growth temperature. As the temperature increases, phase decomposition occurs, transforming the perovskite phase into Ruddlesden-Popper (RP) and NiO phases. (b) High resolution X-ray diffraction (HR-XRD) $2\theta/\omega$ scan data of four NSNO thin films grown on $SrTiO_3$ (STO)(001) substrates at 520 °C (blue), 560 °C (black), 620 °C (brown), and 670 °C (red).

**Figure 2** | Reciprocal space map (RSM) measurements of NSNO thin films were performed around the STO (103) and (013) off-specular peaks. The RSM results demonstrate that the NSNO films are fully strained to the STO substrates.

**Figure 3** | Temperature-dependent resistivity $\rho(T)$ measurements of NSNO thin films. (a) The $\rho(T)$ curves for NSNO1 (blue) and NSNO2 (black) samples. (b) The $\rho(T)$ curves for NSNO3 (blue) and NSNO4 (black) samples. (c) The insulating behavior of NSNO3 and NSNO4 samples, which was well-fitted using the Arrhenius equation.

**Figure 4** | X-ray photoemission spectroscopy (XPS) spectra of Nd $3d_{5/2}$ and Ni $2p_{3/2}$ core levels, along. with fitting results for four NSNO thin films. (a) The Nd $3d_{5/2}$ spectra consist of $Nd^{3+}$ and satellite peaks. (b) The Ni $2p_{3/2}$ spectra consist of $Ni^{2+}$, $Ni^{3+}$, and satellite peaks.

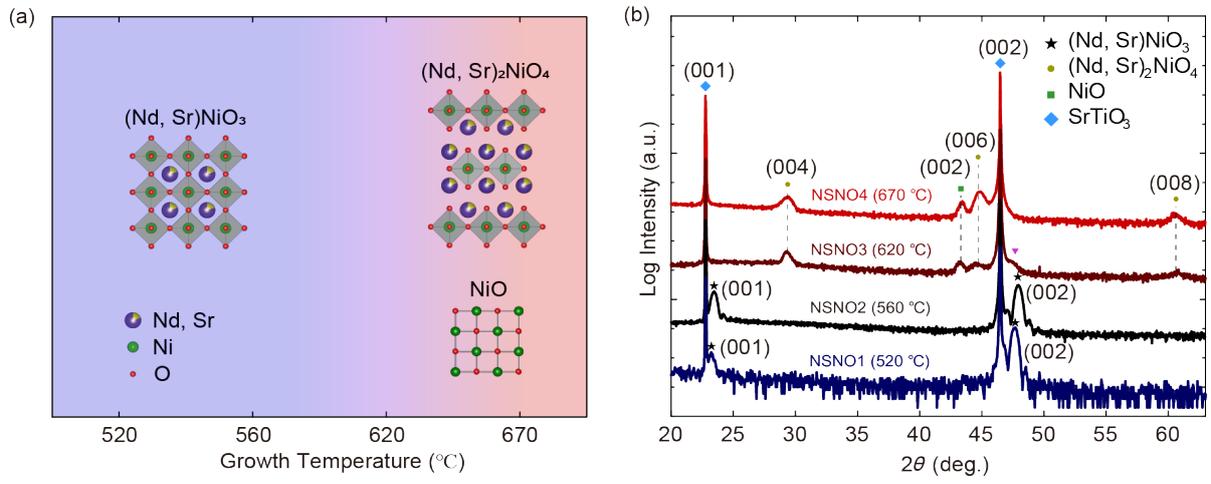

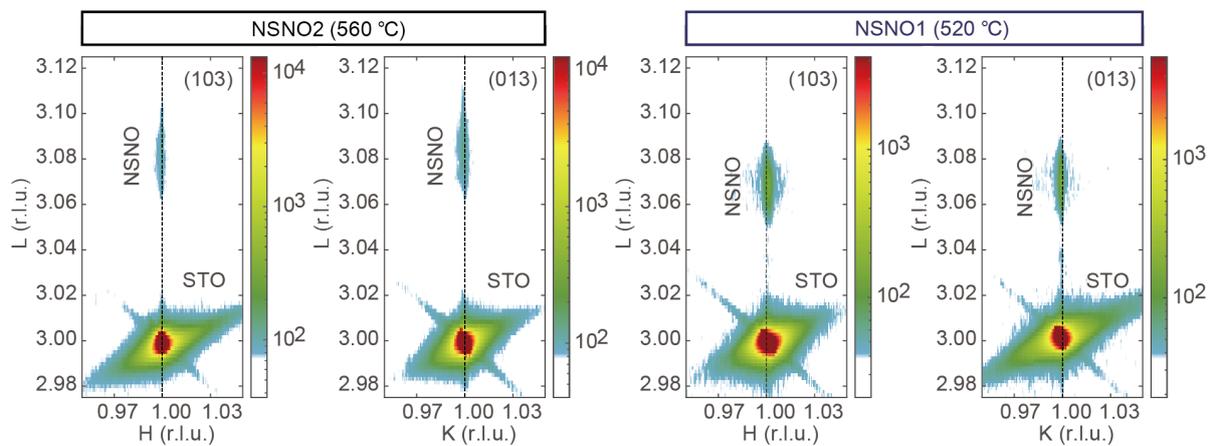

C. Kim *et al*., Fig. 2.

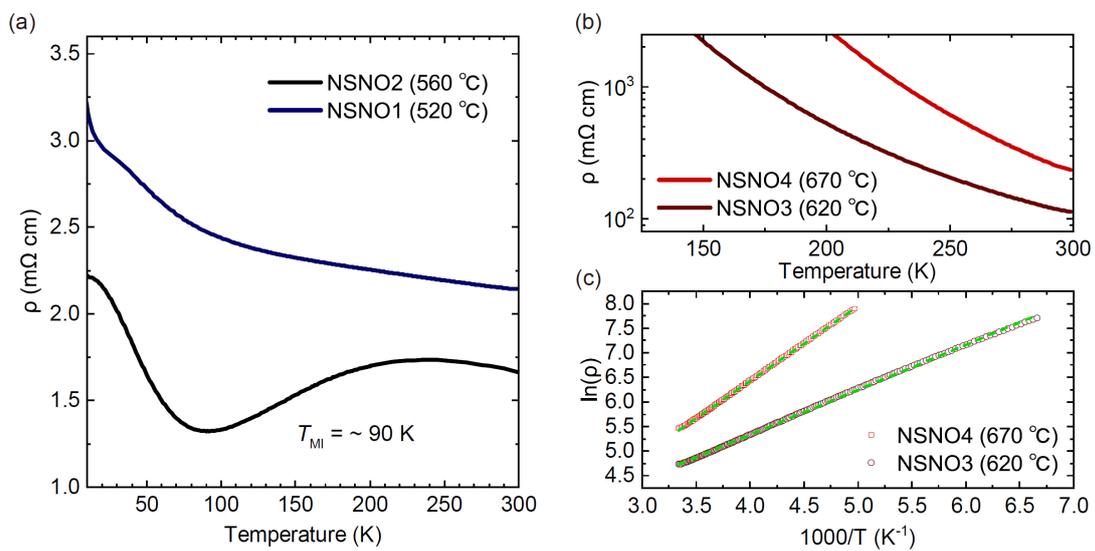

C. Kim *et al.*, Fig. 3.

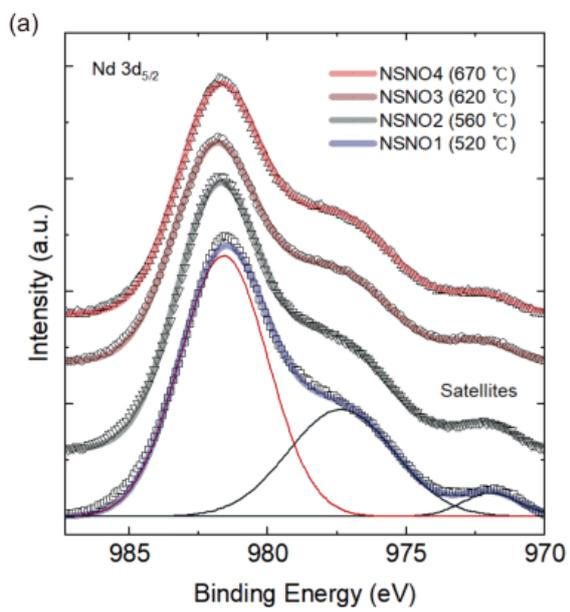 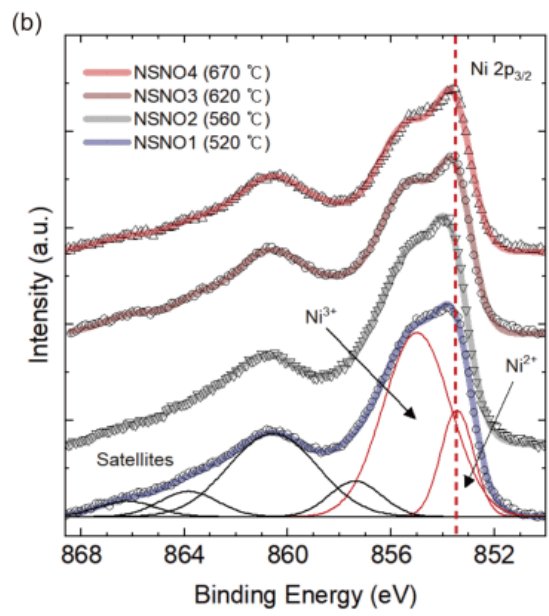

C. Kim *et al*., Fig. 4.